\documentclass[conference]{IEEEtran}

\usepackage{enumerate}
\usepackage{multirow,array}
\usepackage{cite}
\usepackage{graphicx}
\usepackage{psfrag}
\usepackage{caption}
\usepackage{subcaption}
\usepackage{url}
\usepackage{amsmath}
\usepackage{amsthm}
\usepackage{array}
\usepackage{amssymb}
\usepackage{amsfonts}
\usepackage{float}
\usepackage{tabu}

\usepackage{algorithm}
\usepackage{algpseudocode}

\usepackage{color, colortbl}
\definecolor{Gray}{gray}{0.9}

\makeatletter
\renewcommand{\boxed}[1]{\text{\fboxsep=.2em\fbox{\m@th$\displaystyle#1$}}}
\makeatother

\newcommand{\fq}{{\mathbb F}_q}
\newcommand{\gbinom}[2]{\begin{bmatrix}#1\\#2\end{bmatrix}_q}

\newtheorem*{conjecture*}{Conjecture}
\newtheorem{lemma}{Lemma}

\newtheorem{corollary}{Corollary}
\newtheorem{remark}{Remark}
\newtheorem{definition}{Definition}
\newtheorem{theorem}{Theorem}

\title{Cache-Aided Interference Management with Subexponential Subpacketization}
\begin{document}

\author{
\IEEEauthorblockN{Hari Hara Suthan Chittoor, K V Sushena Sree, Prasad Krishnan\\}
\IEEEauthorblockA{
Signal Processing and Communications Research Center,\\
International Institute of Information Technology, Hyderabad, India.\\
Email: \{hari.hara@research., sushena.sree@research., prasad.krishnan@\}iiit.ac.in}
}

\date{\today}
\maketitle
\thispagestyle{empty}	
\pagestyle{empty}
\begin{abstract}
Consider an interference channel consisting of $K_T$ transmitters and $K_R$ receivers with AWGN noise and complex channel gains, and with $N$ files in the system. The one-shot $\mathsf{DoF}$ for this channel is the maximum number of receivers which can be served simultaneously with vanishing probability of error as the $\mathsf{SNR}$ grows large, under a class of schemes known as \textit{one-shot} schemes. Consider that there exists transmitter and receiver side caches which can store fractions $\frac{M_T}{N}$ and $\frac{M_R}{N}$ of the library respectively. Recent work for this cache-aided interference channel setup shows that, using a carefully designed prefetching(caching) phase, and a one-shot coded delivery scheme combined with a proper choice of beamforming coefficients at the transmitters, we can achieve a $\mathsf{DoF}$ of $t_T+t_R$, where $t_T=\frac{M_T K_T}{N}$ and $t_R=\frac{M_R K_R}{N},$ which was shown to be almost optimal. The existing scheme involves splitting the file into $F$ subfiles (the parameter $F$ is called the \textit{subpacketization}), where $F$ can be extremely large (in fact, with constant cache fractions, it becomes exponential in $K_R$, for large $K_R$). In this work, our first contribution is a scheme which achieves the same $\mathsf{DoF}$ of $t_T+t_R$ with a smaller subpacketization than prior schemes. Our second contribution is a new coded caching scheme for the interference channel based on projective geometries over finite fields which achieves a one-shot $\mathsf{DoF}$ of $\Theta(log_qK_R+K_T)$, with a subpacketization $F=q^{O(K_T+(log_qK_R)^2)}$ (for some prime power $q$) that is \textit{subexponential} in $K_R$, for small constant cache fraction at the receivers.  To the best of our knowledge, this is the first coded caching scheme with subpacketization subexponential in the number of receivers for this setting.
\end{abstract}


\section{Introduction}
Modern communication networks have moved beyond traditional point-to-point communication scenarios into multiterminal communication. Many communication scenarios in modern networks also involve delivery of large content \cite{Cis}, with low latency requirements. Owing to the reduction in the cost of storage media, a reasonable option to reduce the network load is  to cache a fraction of the total content near the receivers, and deliver the rest of the requested files through the main network. This technique was formally presented for the single transmitter noiseless broadcast network in \cite{MaN}. It was shown in \cite{MaN} that huge gains can be achieved by a properly jointly designed caching and a delivery scheme which involves sending coded file-parts corresponding to demands of multiple users at the same time. The rate achieved by the scheme in \cite{MaN} was in fact shown to be optimal for uncoded caching in \cite{WTP}. The gains shown using the \textit{coded caching} paradigm established by \cite{MaN} were extended to a number of other scenarios, for instance, D2D communication \cite{D2D}, combination networks \cite{combinationnetworks}, and also multisource-multisink settings as in \cite{interferencemanagement}.

In \cite{interferencemanagement}, the authors consider the problem of managing interference in a multi-source ($K_T$ of them) multi-sink ($K_R$ of them) setting connected through an AWGN channel with complex channel gains between the transmitters and the receivers. Each of the receivers can demand one of the $N$ files present in the library. Each transmitter has some local storage (cache) which can store a $\frac{M_T}{N}$ fraction of the library, while each receiver has a cache that can store a $\frac{M_R}{N}$ fraction. A centralized caching scheme and a one-shot  delivery scheme that employs linear combinations of subfiles, developed based on the scheme in \cite{MaN}, is designed in \cite{interferencemanagement}, which serves $t_T+t_R$ (where $t_T= \frac{{M_T}{K_T}}{N}$ and $t_R=\frac{{M_R}{K_R}}{N}$)  receivers in each transmission round, which is known as the \textit{achievable one-shot linear sum-$\mathsf{DoF}$} (degrees of freedom). The scheme achieves this $\mathsf{DoF}$ using two tools, (a) using the transmitter cache content to zero-force interference at the receivers, and (b) by designing a coded one-shot delivery scheme based on \cite{MaN} exploiting the receiver caches.  This achievable $\mathsf{DoF}$ is then shown to be within a multiplicative factor of two of the optimal one-shot $\mathsf{DoF}$ for the interference channel setting of \cite{interferencemanagement}. 

However, this near-optimality comes at a price. The implementation of the scheme in \cite{interferencemanagement} requires that the number of subfiles of any file, called the \textit{subpacketization}, should be at least $\binom{K_T}{t_T}\binom{K_R}{t_R}\binom{K_R-t_R-1}{t_T-1}t_R!(t_T-1)!$. As $K_R$ grows large (as to be expected in practical scenarios) for constant cache fraction $\frac{M_R}{N}$, the subpacketization grows exponentially in $K_R$ As the subpacketization parameter associated with a coded caching scheme gives a lower bound on the size of the files in order to implement the scheme, this presents a practical difficulty that has to be overcome to reap the benefits of coded caching. The root of this problem lies in the scheme of \cite{MaN} for broadcast settings, and all schemes which utilize this scheme of \cite{MaN} also inherit this problem.


\subsection{Other related Work}
The first work which considered applying the strategy of coded caching to interference channels was \cite{CacheAidedInterferenceChannels}. The authors of \cite{CacheAidedInterferenceChannels} focused on the case with only transmitter caches, which were exploited to get both interference cancellation and interference alignment gains. This was further expanded upon by \cite{DoFHachem} where the DoF region of the interference channel with both transmitter and receiver channels was characterized and a near-optimal scheme (optimal except for a multiplicative gap) was obtained using general delivery schemes (beyond one-shot schemes). Coded caching for multi-antenna interference channels was considered in \cite{MultiAntennaInterference}. A one-shot delivery based coded caching scheme for the setup of \cite{interferencemanagement} was presented in \cite{hypercube_interference}, which achieves smaller subpacketization than the scheme in \cite{interferencemanagement}. However the subpacketization still remains exponential in the number of receivers $K_R$.   
\subsection{Contributions and Organization}
The present work involves low subpacketization constructions of coded caching schemes for the interference management problem under the communication scenario in \cite{interferencemanagement}. The contributions and organization of this work are as follows. 
\begin{itemize}
    \item 
    We first review the system model from \cite{interferencemanagement} and also present the conditions for valid transmission rounds of a one-shot delivery scheme in Section \ref{systemmodel}. By doing this, we setup the strategy for designing our delivery schemes. 
    \item 
    In Section \ref{modified1}, we show that we can achieve the same $\mathsf{DoF}$ as \cite{interferencemanagement}, with a smaller subpacketization of $\binom{K_T}{t_T}\binom{K_R}{t_R}\binom{K_R-t_R-1}{t_T-1}$. However this subpacketization continues to be exponential in $K_R$ for constant $\frac{M_R}{N}.$ We do this by giving a modified delivery scheme of \cite{interferencemanagement} for the caching scheme of \cite{interferencemanagement}. 
    \item Inspired from subexponential subpacketization schemes for the  broadcast coded caching setting developed in \cite{haribhavanaprasad}, in Section \ref{projective geometry based scheme}, we develop a coded caching scheme based on projective geometry over finite fields, which has subpacketization subexponential ($q^{O(K_T+(log_qK_R)^2)}$, where $q$ is some prime power) in the number of receivers $K_R$ for receiver cache fraction $\frac{M_R}{N}$ upper bounded by a small constant (these asymptotics are shown in Section \ref{analysis}). We give a numerical comparison of our projective geometry based scheme with the scheme proposed in Section \ref{modified1} (which is improved from \cite{interferencemanagement}), and also with \cite{hypercube_interference} in Section \ref{projective geometry based scheme} (Table \ref{table 1}). The results show that we outperform all known schemes for the interference channel setup in terms of the subpacketization, however trading it off with a lower achievable $\mathsf{DoF}$ which is $\Theta(log_qK_R+K_T)$. 
\end{itemize}

\textit{Notations and Terminology:} $\mathbb{Z}^{+}$ denotes the set of positive integers. 
We denote the set $\{1,\hdots,n\}$ by $[n]$ for some $n \in \mathbb{Z}^{+}$. For sets $A,B$, the set of elements in $A$ but not in $B$ is denoted by $A\backslash B$. For some element $x$, we denote $A\backslash\{x\}$ by $A\backslash x$ also. The set of $r$ sized subsets of a set $A$ is denoted by $\binom{A}{r}$. 
For $i,j,m\in \mathbb{Z}^{+}$, we define $i\boxplus_{m}j \triangleq  1+((i+j-1) \textit{ mod } m)$. The finite field with $q$ elements is $\fq$. The dimension of a vector space $V$ over $\fq$ is given as $dim(V)$. For two subspaces $V,W$, their subspace sum is denoted by $V+W$. Note that $V+W=V\oplus W$ (the direct sum) if $V\cap W=\phi$. The span of two vectors $\mathbf{v_1},\mathbf{v_2}\in V$, is represented as $span(\mathbf{v_1},\mathbf{v_2})$.

\section{System Model, Basic Terminologies and a Technical Lemma}
\label{systemmodel}
\begin{figure}
\centering
  \includegraphics[width=0.8\linewidth]{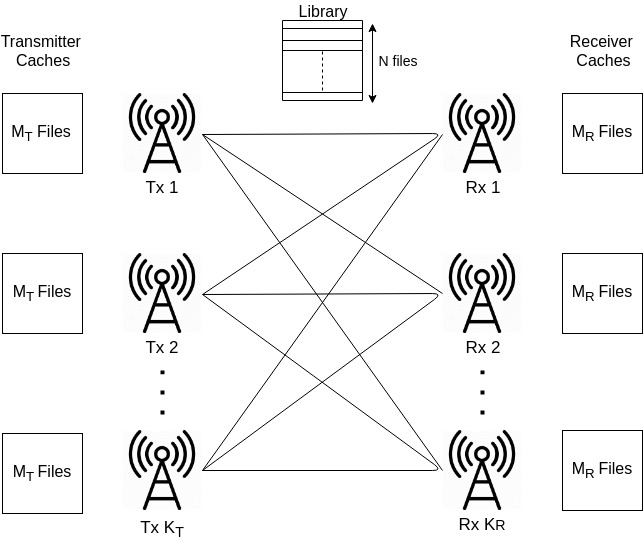}
  \caption{Wireless interference network}
  \label{fig system}
\end{figure}
We follow the model as in \cite{interferencemanagement}. We consider a discrete time AWGN channel with $K_T$ transmitters, denoted by a set ${\cal K}_T$, and $K_R$ receivers, denoted by a set ${\cal K}_R$, as shown in Fig. \ref{fig system}. A collection of $N$ files exist in the system denoted as $W_{n}, n\in[N]$. Each transmitter has the capability of caching a fraction $\frac{M_T}{N}$ of the library, while each receiver can cache a fraction $\frac{M_R}{N}$ of the library. 

The system operates in two phases. In the \textit{caching phase}, each file is divided in $F_C$ subfiles (equal-sized) and prefetched into the caches of the transmitters and the receivers. The subfiles of a file $W_n$ are indexed by a set ${\cal F}_C$ as $W_{n,f}:f\in {\cal F}_C$. In this work, we discuss caching schemes in which the transmitters cache $M_T F_C$ subfiles, the receivers cache $M_R F_C$ subfiles, and also each subfile is cached at $t_T$ transmitters and $t_R$ receivers. Such \textit{symmetric} caching schemes are commonly used in literature. We assume that $\frac{min(M_TK_T,M_RK_R)}{N}\geq 1$, i.e., the cumulative caches at the transmitters (equivalently, the receivers) can hold the entire library.  During the \textit{delivery phase}, each receiver demands one of the files in the library, and the transmitters must cooperatively deliver the missing subfiles of each file to the receivers. In the delivery phase, we allow for further splitting of the subfiles into (equal-sized) \textit{packets} to reduce the delivery time. In this paper, we assume that the number of packets within each subfile of each file is some constant, denoted by $F_P$. 

Designing the coded caching scheme for the interference channel, also known as an \textit{cache-aided interference management scheme}, consists of designing the caching phase (what to place in the cache) and the delivery phase as well. We consider delivery schemes which are \textit{one-shot linear schemes}, as in \cite{interferencemanagement},
described generally as follows. The delivery phase consists of several \textit{rounds} of transmissions. In each round, each transmitter picks some packets corresponding to the subfiles missing in the receiver caches, maps them to complex vectors, and then transmits a linear combination of them. Specifically, let $\boldsymbol{w}_j:j\in [n]$ be some $n$ packets  missing at some receivers, and to be transmitted by some particular transmitter $t\in{\cal T}$. Then the packet $\boldsymbol{w}_j$ is mapped to a complex vector $\boldsymbol{\overline{w}}_j$. Then the transmitter transmits $\boldsymbol{x}_t=\sum_{j=1}^n v_j\boldsymbol{\overline{w}}_j$, where $v_j$ is the complex beamforming coefficients chosen by the transmitter $t$. We assume a power constraint $\frac{1}{b}||\boldsymbol{x}_t||^2\leq \mathsf{SNR}$ on the transmissions, where $b$ denotes the length of $\boldsymbol{x}_t$. 

The output at the receiver $r\in{\cal R}$ is given as
\[
\boldsymbol{y}_r=\sum_{t\in{\cal K}_T}h_{r,t}\boldsymbol{x}_t+\boldsymbol{z}_r,
\]
where $h_{r,t}$ represents the complex channel gain from the transmitter $t$ to the receiver $r$ (assumed to be chosen i.i.d from ${\cal CN}(0,1)$ but remaining constant for a round of transmission), and $\boldsymbol{z}_r$ represents the additive noise vector at receiver $r$, with each component of $\boldsymbol{z}_r$ being i.i.d with distribution ${\cal CN}(0,1)$. 

To precisely define the ability of a transmission round to deliver the packets involved in that round, we define the idea of a \textit{valid transmission round}.
\begin{definition}
For some $n\in{\mathbb Z}^+$, a collection of $n$ packets $P=\{\boldsymbol{w}_r:r=1,..,n\}$ intended to $n$ corresponding distinct receivers $R=\{1,..,n\}$ is said to participate in a \underline{valid transmission round} if there exists a choice of the beamforming coefficients at the transmitters containing $P$ for transmitting a linear combination of $P$ such that each receiver $r$ in $R$ can obtain (after certain interference terms being zero-forced by choice of the beamforming coefficients, and cancelling other interference terms present in cache) a quantity $\overline{\boldsymbol{w}}_r+\boldsymbol{z}_r$ (therefore enabling the decoding of $\boldsymbol{w}_r$ with vanishing error probability as $\mathsf{SNR}$ increases), where $\overline{\boldsymbol{w}}_r$ is the complex vector corresponding to demanded packet $\boldsymbol{w}_r$ and $\boldsymbol{z}_r$ denotes the complex AWGN noise. 
\end{definition}
We then have the definition of \textit{a delivery scheme} as follows.
\begin{definition}
A \underline{(one-shot) delivery scheme} is one in which all the missing packets of the demanded file of each receiver are successfully delivered (i.e., decoded with vanishing error probability with increasing $\mathsf{SNR}$) in $S$ valid transmission rounds (for some finite $S$). The \textit{rate} of the delivery scheme is then defined as 
\begin{align}
\label{rate}
\text{Rate}=\frac{S}{F},
\end{align}
where $F=F_CF_P$ denotes the effective subpacketization (simply, the subpacketization), for $F_C$ being the number of subfiles of each file during the caching phase, and $F_P$ is the number of packets of each subfile during the delivery phase.
\end{definition}



A rate $R$ is said to be \textit{achievable} if there exists a caching scheme and valid delivery scheme with rate $R$. Note that a rate of $K_R(1-\frac{M_R}{N})$ is the naively achievable, with a caching scheme in which $\frac{M_R}{N}$ fraction of each file is cached in the library of each receiver, and a delivery scheme in which the uncached fractions of the receiver demands are delivered on a time-sharing basis. Based on the rate, we now define $\mathsf{DoF}$.

\begin{definition}
[DoF]For an achievable scheme with rate $R,$ we refer to the fraction $\frac{K_R(1-\frac{M_R}{N})}{R}$ as the corresponding \textit{achievable} \textit{one-shot linear $\mathsf{DoF}$ (degrees of freedom)}. The \textit{one-shot linear $\mathsf{DoF}$} of the interference channel is defined as the fraction $\frac{K_R(1-\frac{M_R}{N})}{R^*}$, where $R^*$ is the optimal coded caching rate, i.e., the infimum of all achievable rates given the system parameters $K_T,M_T,K_R,M_R$ and $N$. 
\end{definition}

\begin{remark}
Note that if the delivery scheme serves some $\gamma$ packets to $\gamma$ receivers in each round, then as the total number of missing subfiles is $K_R(1-\frac{M_R}{N})F$, we have from (\ref{rate}) that achievable $\mathsf{DoF}=\gamma$. Thus for such schemes we think of the achievable $\mathsf{DoF}$ as the number of receivers served in each round. 
\end{remark}

This system model is developed in \cite{interferencemanagement}, after which the authors show a construction of a coded caching scheme which achieves a $\mathsf{DoF}$ of $t_T+t_R$ (when $t_T+t_R\leq K_R$), where $t_T=\frac{M_TK_T}{N}, t_R=\frac{M_RK_R}{N}$, by obtaining a caching scheme and delivery scheme inspired from the broadcast coded caching scheme of \cite{MaN}.

In this work, we give new cache-aided interference management schemes with reduced subpacketizations. We now give a small technical lemma, which we shall use in the rest of the paper for this purpose. The proof of the lemma can actually be inferred from \cite{interferencemanagement} (Lemma 3 of \cite{interferencemanagement}), however we give it here in an explicit fashion for our convenience. 

\begin{lemma}
\label{ZF}
Consider a set of $n$ packets $P=\{\boldsymbol{{w}}_{d_{r_1},f_1},...,\boldsymbol{{w}}_{d_{r_n},f_n}\}$, where $\boldsymbol{{w}}_{d_{r_i},f_i}$ is a packet demanded  by receiver $r_i$, and each packet in $P$ is cached at atleast $n-t_T$ receivers among the $n$ receivers $\{r_1,..,r_n\}$ (where $t_T$ refers to number of transmitters in which each packet is cached). Then the set of packets $P$ can participate in a valid transmission round. 
\end{lemma}
\begin{IEEEproof}
On the transmitter side we have, 
\begin{align}
\label{trans}
\begin{bmatrix}
\boldsymbol{x}_1\\ \vdots \\\boldsymbol{x}_{K_T}
\end{bmatrix}=\begin{bmatrix}
\overline{M}_1 & \cdots & \overline{M}_n
\end{bmatrix}\begin{bmatrix}
\boldsymbol{\overline{w}}_{d_{r_1},f_1}\\ \vdots \\\boldsymbol{\overline{w}}_{d_{r_n},f_n}
\end{bmatrix}  
\end{align}
where $\boldsymbol{x}_i \in \mathbb{C}^{1 \times b}: i \in [K_T]$ denotes the signal transmitted by $i^{th}$ transmitter and $\overline{M}_i \in \mathbb{C}^{K_T \times 1} : i \in [n]$ is a column vector of size $K_T \times1$ which denotes the beamforming vector of (complexified) packet $\overline{\boldsymbol{{w}}}_{d_{r_i},f_i}: i \in [n]$. Note that the rows of $\overline{M}_i$ corresponding to the indices of transmitters where packet $\boldsymbol{{w}}_{d_{r_i},f_i}$ is not cached contain zeros. Thus $\overline{M}_i$ can have at most $t_T$ non-zeros for each $i \in [N]$, which are to be chosen by the respective transmitters. We want to show that there exists a choice of beamforming coefficients in $\overline{M}_i, \forall i$ such that packets $\boldsymbol{{w}}_{d_{r_1},f_1},...,\boldsymbol{{w}}_{d_{r_n},f_n}$ can be decoded at corresponding receivers $r_1,...,r_n$.

On the receiver side we have, 

\begin{align}
\label{recv}
    \begin{bmatrix}
\boldsymbol{y}_{r_1}\\ \vdots \\\boldsymbol{y}_{r_n}
\end{bmatrix}=H\begin{bmatrix}
\boldsymbol{x}_1\\ \vdots \\\boldsymbol{x}_{K_T}
\end{bmatrix}+Z
\end{align}

where $\boldsymbol{y}_{r_i}: i \in [n]$ denotes the signal received by receiver $r_i$, $H_{n\times K_T}$ denotes the channel state matrix containing the complex channel gain coefficients and $Z_{n \times b}$ denotes the additive Gaussian noise. 

From (\ref{trans}) and (\ref{recv}) we have, 
\begin{align*}
    \begin{bmatrix}
\boldsymbol{y}_{r_1}\\ \vdots \\\boldsymbol{y}_{r_n}
\end{bmatrix}&=H\begin{bmatrix}
\overline{M}_1 & \cdots & \overline{M}_n
\end{bmatrix}\begin{bmatrix}
\overline{\boldsymbol{{w}}}_{d_{r_1},f_1}\\ \vdots \\\overline{\boldsymbol{{w}}}_{d_{r_n},f_n}
\end{bmatrix}+Z\\
&=C\begin{bmatrix}
\overline{\boldsymbol{{w}}}_{d_{r_1},f_1}\\ \vdots \\\overline{\boldsymbol{{w}}}_{d_{r_n},f_n},
\end{bmatrix}+Z
\end{align*}
where $C=[C_1 \hdots C_n],$ such that $C_i=H \overline{M}_i: i \in [n]$. Let $[v_{k_1} \hdots v_{k_n}]$ denote the $k^{th}$ row of $C$. For successful decoding of $\boldsymbol{{w}}_{d_{r_k},f_k}$ at receiver $r_k$, we should have 
\[
\boldsymbol{y}_{r_k}=\sum_{i=1}^{n}v_{k_i} \overline{\boldsymbol{{w}}}_{d_{r_i},f_i}+\boldsymbol{z}_{r_i},
\]
such that $v_{k_i}=0$, for all $i$ such that $\boldsymbol{{w}}_{d_{r_i},f_i}$ is not present in cache of receiver $r_k$ and $v_{k_i} \neq 0$ for $i=k$ (w.l.o.g, we may assume $v_{k_k}=1$). The beamforming vectors $\overline{M}_i$s have to be chosen such that the above condition holds. Since it is given that any packet $\boldsymbol{{w}}_{d_{r_i},f_i}$ is cached at atleast $n-t_T$ receivers and demanded by $r_i$, the vector $\overline{M}_i$ must be chosen so that the $i^{th}$ column $C_i$ must have precisely $1$ at the $i^{th}$ position and $0$'s in $e \leq (t_T-1)$ positions (corresponding to the row indices of those receivers which have not cached $\boldsymbol{{w}}_{d_{r_i},f_i}$). Let $C'_i$ be the subvector of $C_i$ restricted only to these $e+1$ positions. Then we have,
\[
C'_i = H' \overline{M}_i
\]
where $H'_{(e+1)\times K_T}$ is some appropriate submatrix of $H$. Further since any packet is available at $t_T$ transmitters by the caching design, the packet $\boldsymbol{{w}}_{d_{r_i},f_i}$ is cached in $t_T$ transmitters, which means $K_T-t_T$ entries of $\overline{M}_i$ are zero. Let ${\overline{M^{''}_i}}$ be the subvector of $\overline{M}_i$ after removing all these $K_T-t_T$ zero positions of $\overline{M}_i$. Then 
\[
C'_i = {H^{''}} {\overline{M^{''}_i}},
\]
for some appropriate submatrix $H^{''}_{(e+1)\times t_T}$ of $H'$. Since $e+1 \leq t_T$ and as the channel coefficients are generated i.i.d from ${\cal CN}(0,1)$, the matrix $H^{''}$ is of rank $e+1$. Thus there exists some solution to the vector ${\overline{M^{''}_i}}$, which gives us the beamforming vector $\overline{M}_i$ (by appending zero to ${\overline{M^{''}_i}}$).
\end{IEEEproof}
\begin{remark}
Though we have considered $K_T$ transmitters participating in each valid round of transmission in Lemma \ref{ZF}, it is sufficient to have $t_T$ transmitters participating in each valid round of transmission as used in the constructions in the remainder of the paper.
\end{remark}





\section{A Cache-Aided Interference Management Scheme with Reduced Subpacketization achieving $\mathsf{DoF}=t_T+t_R$}
\label{modified1}

 The work \cite{interferencemanagement} presents a construction of a coded caching scheme for the interference channel achieving $\textsf{DoF}=t_T+t_R=\frac{K_TM_T+K_RM_R}{N}$ (we restrict ourselves to the case $K_R\geq t_T+t_R$) with the subpacketization $F=\binom{K_T}{t_T}\binom{K_R}{t_R}\binom{K_R-t_R-1}{t_T-1}t_R!(t_T-1)!$. In this section, we obtain a new delivery scheme for the same caching scheme which obtains the same $\mathsf{DoF}=t_T+t_R$ with a smaller subpacketization $F=\binom{K_T}{t_T}\binom{K_R}{t_R}\binom{K_R-t_R-1}{t_T-1}$. Note that as a result, the subpacketization of this new scheme can be exponentially smaller than that of \cite{interferencemanagement}, for large $K_R, K_T$. A numerical comparison of these is provided in Table \ref{table 1}, in the end of Section \ref{projective geometry based scheme}. 
\subsubsection{Caching Scheme}
\label{modified_caching}
We use the caching technique given in \cite{interferencemanagement}. To describe the caching strategy, we first breakdown the files present in the library into subfiles. We consider $t_T = \frac{{K_T}{M_T}}{N}$ and $t_R = \frac{{K_R}{M_R}}{N}$. Each subfile $W_i:i \in [N]$ in the library is divided into disjoint subfiles denoted as $\{W_{i,{\cal T},{\cal R}}: {\cal T} \in \binom{[K_T]}{t_T},  {\cal R} \in \binom{[K_R]}{t_R}\}$. Hence the subpacketization during the caching phase is $F_C=\binom{K_T}{t_T}\binom{K_R}{t_R}$. Each transmitter and receiver caches a set of subfiles from the library. The caching strategy on the transmitter side is such that transmitter $t_j \in [K_T]$ caches subfiles $\{W_{i,{\cal T},{\cal R}}: t_j \in {\cal T}\}$. Similarly, the caching strategy on the receiver side is such that receiver $r_j \in [K_R]$ caches subfiles $\{W_{i,{\cal T},{\cal R}}: r_j \in {\cal R}\}$. 
\subsubsection{Transmission Scheme}
\label{modified_transmission}
Let $W_{d_{r_j}}$ denote the file demanded by $r_j\in [K_R]$ in the delivery phase. Thus, each receiver $r_j \in [K_R]$ is to be served the subfiles $\{W_{d_{r_j},{\cal T},{\cal R}}: r_j \notin {\cal R},d_{r_j} \in [N]\}$ which are not available in its cache.

We seek to construct valid transmission rounds following Lemma \ref{ZF}. Towards that end, we further divide the demanded subfiles into packets and transmit $t_T+t_R$ packets in each round of transmission. Every subfile $W_{d_{r_j},{\cal T},{\cal R}}$ demanded by receiver $r_j$ is divided into $\binom{K_R-t_R-1}{t_T-1}$ packets denoted as 
\[
    W_{d_{r_j},{\cal T},{\cal R}}= \bigg\{W_{d_{r_j},{\cal T},{\cal R},{\cal R'}}: {\cal R'} \in \binom{[K_R]\setminus{({\cal R} \cup \{r_j\})}}{t_T-1}\bigg\}
\]
The index ${\cal R'}$ essentially denotes the indices of receivers at which the packet $W_{d_{r_j},{\cal T},{\cal R},{\cal R'}}$ is zero-forced. 

We now describe the set of packets that can be sent in a valid round of transmission. For a set ${\cal U} \in \binom{[K_R]}{t_T+t_R}$ denoted as $\{u_1,...,u_{t_T+t_R}\} $, consider some set  $U \subset {\cal U}$ denoted as $\{u_{i_1},\cdots,u_{i_{t_R}}\}$, then we define $U\boxplus_{|\cal U|} l \triangleq \{u_{(i_1\boxplus_{|\cal U|} l)},\cdots,u_{(i_n\boxplus_{|\cal U|} l)}\}$ where $l \in \{0,..,t_T+t_R-1\}$.
\begin{lemma}
\label{transmission lemma}
For a set ${\cal T} \in \binom{[K_T]}{t_T}$, consider a set ${\cal U} \in \binom{[K_R]}{t_T+t_R}$ denoted as $\{u_1,...,u_{t_T+t_R}\}$. Pick some $u_j \in {\cal U}$. Let $\mathfrak{U}=\binom{{\cal U}\setminus u_j}{t_R}$. Now choose $U \in \mathfrak{U}$. Consider a set of packets ${\cal P}({\cal T},u_j,{\cal U},U)=\{W_{d_{u_{(j {\boxplus}_{|{\cal U}|} l)}},{\cal T},{({U {\boxplus}_{|{\cal U}|} l}),{\cal U}\setminus \big({(U {\boxplus}_{|{\cal U}|} l ) \cup \{u_{(j {\boxplus}_{|{\cal U}|} l)}\} }\big)}}: l \in \{0,..,t_T+t_R-1\}\}$. Then there is a valid round of transmissions in which ${\cal P}({\cal T},u_j,{\cal U},U)$ can participate in. 
\end{lemma} 
\begin{IEEEproof}
The set ${\cal P}({\cal T},u_j,{\cal U},U)$ contains $t_T+t_R$ packets. Consider a packet $W_{d_{u_{(j {\boxplus}_{|{\cal U}|} l)}},{\cal T},{({U {\boxplus}_{|{\cal U}|} l}),{\cal U}\setminus \big({(U {\boxplus}_{|{\cal U}|} l ) \cup \{u_{(j {\boxplus}_{|{\cal U}|} l)}\} }\big)}}$ demanded by receiver $u_{(j {\boxplus}_{|{\cal U}|} l)}$. Following the defined caching scheme, it is easy to check that the packet is cached at the set of $t_R$ receivers $({U {\boxplus}_{|{\cal U}|} l})$. Since each packet in the set ${\cal P}$ is cached at $t_R$ receivers, there exists a valid round of transmission which delivers these packets $\{W_{d_{u_{(j {\boxplus}_{|{\cal U}|} l)}},{\cal T},{({U {\boxplus}_{|{\cal U}|} l}),{\cal U}\setminus \big({(U {\boxplus}_{|{\cal U}|} l ) \cup \{u_{(j {\boxplus}_{|{\cal U}|} l)}\} }\big)}}: l \in \{0,..,t_T+t_R-1\}\}$ to their respective receivers $\{u_{(j {\boxplus}_{|{\cal U}|} l)} : l \in \{0,..,t_T+t_R-1\} \}$ as in Lemma \ref{ZF}.
\end{IEEEproof}
From the above lemma, it is easy to see that any arbitrary packet $W_{d_{u_j},{\cal T},{\cal R},{\cal R'}}$ can be decoded from the unique round of transmission consisting of the set of packets ${\cal P}\big({\cal T},u_j,(u_j\cup{\cal R} \cup {\cal R'}),{\cal R}\big)$. We thus have Algorithm \ref{trans algorithm} which describes the transmission algorithm.
\begin{algorithm}[htbp]
\caption{Transmission Scheme}
\label{trans algorithm}
\begin{algorithmic}[1]
\Procedure{Transmissions}{}
    \For{each ${\cal T} \in \binom{[K_T]}{t_T}$}
    \For {each ${\cal U} \in \binom{[K_R]}{t_T+t_R}$}
    \State pick $u_j \in {\cal U}$ and let $\mathfrak{U}=\binom{{\cal U}\setminus{u_j}}{t_R}$
    \For {each $U \in \mathfrak{U}$}
    \State Obtain a valid round of transmission (as in Lemma \ref{ZF}) using the packets $\{W_{d_{u_{(j {\boxplus}_{|{\cal U}|} l)}},{\cal T},{({U {\boxplus}_{|{\cal U}|} l}),{\cal U}\setminus \big({(U {\boxplus}_{|{\cal U}|} l ) \cup \{u_{(j {\boxplus}_{|{\cal U}|} l)}\} }\big)}}: l \in \{0,..,t_T+t_R-1\}\}$
    \EndFor
    \EndFor
    \EndFor
\EndProcedure
\end{algorithmic}
\end{algorithm}

We thus have the theorem which summarizes our results.
\begin{theorem}
Given a caching scheme as described in Section \ref{modified_caching}, we get a transmission scheme with subpacketization level $F=\binom{K_T}{t_T}\binom{K_R}{t_R}\binom{K_R-t_R-1}{t_T-1}$ and $\mathsf{DoF}= t_T+t_R$.
\end{theorem}


\section{A new projective geometry based scheme with low subpacketization}
\label{projective geometry based scheme}
In this section, we obtain a projective geometry based scheme which achieves subexponential subpacketization, however at the cost of lower $\mathsf{DoF}$. In particular, we achieve a $\mathsf{DoF}$ of $\Theta(log_qK_R+K_T)$, with a subpacketization $F=q^{O(K_T+(log_qK_R)^2)}$ for constant cache fractions at transmitters and receivers ($\frac{M_T}{N}\leq \frac{1}{q^{\alpha-1}},\frac{M_R}{N}\leq \frac{1}{q^{\beta-1}}$, for $\alpha,\beta\in{\mathbb Z}^+$) .  Towards presenting our new scheme, we first review some basic concepts from projective geometry.
\vspace{-0.3cm}
\subsection{Review of projective geometries over finite fields \cite{hirschfeld1998projective}}
Let $a,b,q\in \mathbb{Z}^+$ such that $q$ is a prime power. Let $\fq^{a}$ be $a$-dim vector spaces over a finite field $\fq$. Consider an equivalence relation on $\fq^a\setminus\{\boldsymbol{0}\}$ (where $\boldsymbol{0}$ represents the zero vector) whose equivalence classes are $1$-dim subspaces (without $\boldsymbol{0}
$) of $\fq^a$. The $(a-1)$-dim \textit{projective space} over $\fq$ is denoted by $PG_q(a-1)$ and is defined as the set of these equivalence classes. 
For $b\in [a]$, let $PG_q(a-1,b-1)$ denote the set of all $b$-dim subspaces of $\fq^{a}$.
From Chapter $3$ in \cite{hirschfeld1998projective}. it is known that $|PG_q(a-1,b-1)|$ is equal to the \textit{Gaussian-binomial coefficient} $\gbinom{a}{b}$, where
$
\begin{bmatrix}a\\b\end{bmatrix}_q
=\frac{(q^{a}-1)\hdots(q^{a-b+1}-1)}{(q^{b}-1)\hdots(q-1)}
$ (where $a\geq b$).
In fact, $\gbinom{a}{b}$ gives the number of $b$-dim subspaces of any $a$-dim vector space over $\fq$. Further, by definition, $\gbinom{a}{0}=1.$

Let $\mathbb{C} \triangleq \{C : C \in PG_q(a-1,0) \}$. Let $\theta(a)$ denotes the number of distinct $1$-dim subspaces of $\mathbb{F}_q^a$. Therefore $\theta (a) = |\mathbb{C}|=\gbinom{a}{1}= \frac{q^a -1}{q-1}$.  

The following lemma and corollary of \cite{hari_SPAWC} will be used to prove the statements in forth coming subsections.
\begin{lemma} \cite{hari_SPAWC} \label{no of sets of LI 1D spaces}
Let $k,a,b \in \mathbb{Z}^+$ such that $1\leq a+b\leq k$. Consider a $k$-dim vector space $V$ over $\fq$ and a fixed $a$-dim subspace $A$ of $V$. The number of distinct (un-ordered) $b$-sized sets $\{C_1,C_2,\cdots,C_b\}$ such that $C_i\in \mathbb{C}, \forall i\in[b]$ and $(A\oplus C_1 \oplus C_2 \oplus \cdots \oplus C_b) \in PG_q(k-1,a+b-1)$ is 
$\frac{\prod\limits_{i=0}^{b-1}(\theta(k)-\theta(a+i))}{b!}$.
\end{lemma}

\begin{corollary} \cite{hari_SPAWC} \label{no of 1D spaces outside a hyper space}
Consider two subspaces $A,A'$ of a $k$-dim vector space $V$ over $\fq$ such that $A'\subseteq A, dim(A)=a, dim(A')=a-1$. The number of distinct $C\in \mathbb{C}$ such that $A'\oplus C=A$ is $q^{a-1}$.
\end{corollary}

We now proceed to construct  a new coded caching scheme for interference channel using projective geometry.
\subsection{Projective geometry based caching scheme} \label{our scheme B subsection}
Consider $k_t,m_t,l_t,k_r,m_r,l_r,q \in \mathbb{Z}^+$ such that $q$ is a prime power, $k_t\geq m_t+l_t$ and $k_r\geq m_r+l_r+\gbinom{m_t+1}{1}$. Consider $k_t$-dim vector space $\fq^{k_t}$ and $k_r$-dim vector space $\fq^{k_r}$. Let
\begin{align*}
\mathbb{C}_t \triangleq \{C : C \in PG_q(k_t-1,0) \}. \\
\mathbb{C}_r \triangleq \{C : C \in PG_q(k_r-1,0) \}.    
\end{align*}

Therefore $\mathbb{C}_t,\mathbb{C}_r$ represents set of all distinct $1$-dim subspaces of $\fq^{k_t}$ and $\fq^{k_r}$ respectively.
Let $L_t$ be a fixed $(l_t-1)$-dim subspace of $\fq^{k_t}$. Consider the following sets of subspaces of $\fq^{k_t}$, where each such subspace contains $L_t$.
\begin{align*}
    \mathbb{U} &\triangleq \{U \in PG_q(k_t-1,l_t-1): L_t\subseteq U \}. \text{(indexes Tx's)}.\\
    \mathbb{P}_t &\triangleq \{P_t \in PG_q(k_t-1,m_t+l_t-1): L_t\subseteq P_t \}.\\
    \mathbb{X}_t & \triangleq \left\{\{U_1,U_2,\cdots,U_{m_t+1}\}: \forall U_i \in \mathbb{U}, \sum\limits_{i=1}^{m_t+1}U_i \in \mathbb{P}_t \right\}.
\end{align*}

Let $L_r$ be a fixed $(l_r-1)$-dim subspace of $\fq^{k_r}$. Consider the following sets of subspaces of $\fq^{k_r}$, where each such subspace contains $L_r$.
\begin{align*}
    \mathbb{V} &\triangleq \{V \in PG_q(k_r-1,l_r-1): L_r\subseteq V \}.\text{(indexes Rx's)}. \\
    \mathbb{P}_r &\triangleq \{P_r \in PG_q(k_r-1,m_r+l_r-1): L_r\subseteq P_r \}.\\
    \mathbb{X}_r & \triangleq \left\{\{V_1,V_2,\cdots,V_{m_r+1}\}: \forall V_i \in \mathbb{V}, \sum\limits_{i=1}^{m_r+1}V_i \in \mathbb{P}_r \right\}.
\end{align*}

Let $\{W_i, \forall i\in [N]\}$ be the set of $N$ files available at the library. Each transmitter has capacity to store $M_T$ number of files and each receiver has capacity to store $M_R$ number of files.

Let $\mathbb{U},\mathbb{V}$ denote the set of transmitters and receivers respectively. Therefore $K_T=|\mathbb{U}|$ and $K_R=|\mathbb{V}|$. During the caching phase files are divided into subfiles which are represented as $W_{i,X_t,X_r}: i\in [N], X_t \in \mathbb{X}_t, X_r\in \mathbb{X}_r$.
Let $F_T=|\mathbb{X}_t|$, $F_R=|\mathbb{X}_r|$. Thus $F_C=F_T F_R$ is the subpacketization in the caching phase.



We now present an algorithm for our projective geometry based caching scheme in Algorithm \ref{caching algorithm}.


\begin{algorithm}
\caption{Caching scheme}
\label{caching algorithm}
\begin{algorithmic}[1]
\Procedure{Placement Phase}{}
    \For{each $i\in [N]$}
    \State Split $W_i$ into $\{W_{i,X_t,X_r}:X_t\in\mathbb{X}_t,X_r\in\mathbb{X}_r\}$.
    \EndFor
    \For {each $U \in \mathbb{U}$}
    \State Transmitter $U$ caches the subfiles $W_{i,X_t,X_r}, \forall i\in [N], \forall X_r \in \mathbb{X}_r$ and $ \forall X_t\in \mathbb{X}_t$ such that $U\subseteq \sum\limits_{U_i \in X_t}U_i$.
    \EndFor
    \For {each $V \in \mathbb{V}$}
    \State Receiver $V$ caches the subfiles $W_{i,X_t,X_r}, \forall i\in [N], \forall X_t \in \mathbb{X}_t$ and $ \forall X_r\in \mathbb{X}_r$ such that $V\subseteq \sum\limits_{V_i \in X_r}V_i$.
    \EndFor
\EndProcedure
\end{algorithmic}
\end{algorithm}

We now find the values of $K_T,K_R,F_T,F_R,t_T,t_R$.
Consider an arbitrary subfile $W_{i,X_t,X_r}$.
Let $t_T,t_R$ represents the number of transmitters and receivers this subfile is cached respectively.
It is easy to see that,
$t_T=|\{U:U \subseteq \sum\limits_{U_i\in X_t}U_i \}|$ and $t_R=|\{V:V \subseteq \sum\limits_{V_i\in X_r}V_i \}|$.

\begin{lemma}
\label{K_T,K_R,t_T,t_R,F expressions}
\begin{itemize}
    \item $K_T =|\mathbb{U}| = \gbinom{k_t-l_t+1}{1}.$ 
    \item $K_R =|\mathbb{V}| = \gbinom{k_r-l_r+1}{1}.$
    \item $F_C=F_T F_R$  \text{,  where}
    \begin{itemize}
        \item $F_T=|\mathbb{X}_t| = \frac{q^{\frac{m_t(m_t+1)}{2}}}{(m_t+1)!}\prod\limits_{i=0}^{m_t}\gbinom{k_t-l_t+1-i}{1}.$
        \item $F_R=|\mathbb{X}_r| = \frac{q^{\frac{m_r(m_r+1)}{2}}}{(m_r+1)!}\prod\limits_{i=0}^{m_r}\gbinom{k_r-l_r+1-i}{1}.$
    \end{itemize}
    \item $t_T = \gbinom{m_t+1}{1}.$
    \item $t_R = \gbinom{m_r+1}{1}.$
\end{itemize}

\end{lemma}

\begin{IEEEproof}
Finding $K_T=|\mathbb{U}|$: 
Finding $|\mathbb{U}|$ is equivalent to counting the number of distinct $ C \in \mathbb{C}_t$ which gives distinct $L_t \oplus C \in \mathbb{U}$.
By Lemma \ref{no of sets of LI 1D spaces} and Corollary \ref{no of 1D spaces outside a hyper space} we have, 
$K_T=\frac{\theta(k_t)-\theta(l_t-1)}{q^{l_t-1}} = \frac{q^{k_t}-q^{l_t-1}}{q^{l_t-1}(q-1)}=\frac{q^{k_t-l_t+1}-1}{q-1}= \gbinom{k_t-l_t+1}{1}$. 

Similarly we can prove $K_R$.

Finding $|\mathbb{X}_t|$: 
Finding $|\mathbb{X}_t|$ is equivalent to counting the number of distinct sets $ \{C_1,C_2,\cdots, C_{m_t+1}\}$ (such that $ C_i\in \mathbb{C}_t~\forall i\in [m_t+1]$ and $L_t\oplus C_1 \oplus C_2 \oplus \cdots \oplus C_{m_t+1} \in  \mathbb{P}_t $) which gives distinct $\{L_t\oplus C_1,L_t\oplus C_2, \cdots ,L_t\oplus C_{m_t+1}\} \in \mathbb{X}_t$.
By Lemma \ref{no of sets of LI 1D spaces} we have, the number of distinct sets $ \{C_1,C_2,\cdots,C_{m+1}\}$, such that $ C_i\in \mathbb{C}_t ~(\forall i\in [m_t+1])$ and $L_t\oplus C_1 \oplus C_2 \cdots \oplus C_{m_t+1} \in  \mathbb{P}_t $, is 
$\frac{\prod\limits_{i=0}^{m_t}(\theta(k_t)-\theta(l_t-1+i))}{(m_t+1)!}$. It is easy to check that $\{L_t\oplus C_1,L_t\oplus C_2, \cdots ,L_t\oplus C_{m_t+1}\} \in \mathbb{X}_t$. By Corollary \ref{no of 1D spaces outside a hyper space} we have, the number of distinct $C\in \mathbb{C}_t$ such that $L_t\oplus C= U$ for some fixed $U\in \mathbb{U}$ is $q^{l_t-1}$. Therefore for each $\{L_t\oplus C_1,L_t\oplus C_2, \cdots ,L_t\oplus C_{m_t+1}\} \in \mathbb{X}_t$ there exist $(q^{l_t-1})^{m_t+1}=q^{(m_t+1)(l_t-1)}$ distinct $\{C_1',C_2',\cdots, C_{m_t+1}'\}$ (where $C_i'\in \mathbb{C}_t~ \forall i\in [m_t+1]$) such that $L_t\oplus C_i=L_t\oplus C_i', \forall i\in [m_t+1]$. 
Therefore we can write 
\begin{align*}
 |\mathbb{X}_t|&=\frac{\prod\limits_{i=0}^{m_t}(\theta(k_t)-\theta(l_t-1+i))}{(m_t+1)!~ q^{(m_t+1)(l_t-1)}}\\
&= \frac{\prod\limits_{i=0}^{m_t}(q^{k_t}-q^{l_t-1+i})}{(m_t+1)!~ q^{(m_t+1)(l_t-1)}~(q-1)^{(m_t+1)}}\\
&= \frac{q^{(m_t+1)(l_t-1)}\left(\prod\limits_{i=0}^{m_t}q^i\right)\left(\prod\limits_{i=0}^{m_t}(q^{k_t-l_t+1-i}-1)\right)}{(m_t+1)!~ q^{(m_t+1)(l_t-1)}~(q-1)^{(m_t+1)}} \\
&= \frac{q^{\frac{m_t(m_t+1)}{2}}}{(m_t+1)!}\prod\limits_{i=0}^{m_t}\gbinom{k_t-l_t+1-i}{1}.
\end{align*}
Similarly we can prove $|\mathbb{X}_r|$.

Finding $t_T$: Consider an arbitrary subfile $W_{i,X_t,X_r}$. We have
$t_T=|\{U:U \subseteq \sum\limits_{U_i\in X_t}U_i \}|$.
We know that $\sum\limits_{U_i\in X_t}U_i =P_t$ for some $P_t\in \mathbb{P}_t$. Now finding $t_T$ is equivalent to counting the number of distinct $C\in \mathbb{C}_t$, which gives distinct $L_t \oplus C \in \mathbb{U}$ such that $L_t \oplus C\subseteq P_t$. By Lemma \ref{no of sets of LI 1D spaces} and Corollary \ref{no of 1D spaces outside a hyper space} we have, $t_T=\frac{\theta(m_t+l_t)-\theta(l_t-1)}{q^{(l_t-1)}} = \frac{q^{(m_t+l_t)}-q^{(l_t-1)}}{q^{(l_t-1)}(q-1)}= \frac{q^{(m_t+1)}-1}{q-1}=\gbinom{m_t+1}{1} $. Similarly we can prove $t_R$.
\end{IEEEproof}


\subsection{Projective geometry based delivery scheme}
We now present our delivery scheme for the caching scheme presented above. This delivery scheme is essentially inspired from Section \ref{modified1}. We first split the demanded subfiles into packets, and then transmit the missing packets by coding them across multiple rounds of transmissions, each of which are obtained according to Lemma \ref{ZF}. Towards this end, we consider the following sets.

$\mathbb{D} \triangleq \{D \in PG_q(k_r-1,l_r+t_T-3): L_r\subseteq D \}.$

$\mathbb{Y} \triangleq \left\{\{V_1,V_2,\cdots,V_{t_T-1}\}: \forall V_i \in \mathbb{V}, \sum\limits_{i=1}^{t_T-1}V_i \in \mathbb{D} \right\}.$

$\mathbb{E} \triangleq \{E \in PG_q(k_r-1,m_r+l_r+t_T-1): L_r\subseteq E \}.$

$\mathbb{Z} \triangleq \left\{\{V_1,\cdots,V_{m_r+t_T+1}\}: \forall V_i \in \mathbb{V}, \sum\limits_{i=1}^{m_r+t_T+1}V_i \in \mathbb{E} \right\}.$ 

Note that for an arbitrary demanded subfile $W_{d_{V_a},X_t,X_r}$ for some $X_r\in \mathbb{X}_r$, we have that $(V_a+\sum\limits_{V_i\in X_r}V_i)$ is a $(m_r+l_r+1)$-dim subspace containing $L_r$. We now present the subfile-splitting technique and the delivery scheme in Algorithm \ref{transmission algorithm}. For the purpose of the key step (Step 11) of the Algorithm, we need to define a notation. For some $Z=\{V_1,\hdots,V_{m_r+t_T+1}\}\in{\mathbb Z}$, let $S=\{V_{i_1},...,V_{i_{m_r+1}}\}\subset Z$. Then, for any $l\in\{0,1,\hdots,m_r+t_T\}$, define 
\[
S\boxplus_{|Z|}l\triangleq\{V_{(i_1\boxplus_{|Z|}l)},...,V_{(i_{m_r+1}\boxplus_{|Z|}l)}\}.
\]



\begin{algorithm}
\caption{Transmission scheme}
\label{transmission algorithm}
\begin{algorithmic}[1]
\Procedure{Splitting of demanded subfiles}{demand of receiver $V$ is represented as $W_{d_{V}}$}
    \For{each $V\in{\mathbb V}$, each $X_t\in \mathbb{X}_t$ and each $X_r\in{\mathbb X}_r$}
    \State Split $W_{d_V,X_t,X_r}$ into $\{W_{d_V,X_t,X_r,Y}:Y\in\mathbb{Y}$ and $\{V\}\cup X_r\cup Y\in \mathbb{Z}\}$.
    \EndFor
\EndProcedure
\Procedure{Transmissions}{}
    \For{each $X_t \in \mathbb{X}_t$}
    \For {each $Z \in \mathbb{Z}$}
    \State pick $V_j \in \mathbb{Z}$ and let $\mathfrak{S}=\binom{Z\setminus{V_j}}{m_r+1}$
    \For {each $S \in \mathfrak{S}$}
    \State Obtain a valid round of transmission (as in Lemma \ref{ZF}) using the packets ${\cal P}(X_t,V_j,Z,S)=\{W_{d_{V_{j {\boxplus}_{|Z|} l}},X_t,{({S {\boxplus}_{|Z|} l}),Z\setminus \big({(S {\boxplus}_{|Z|} l ) \cup \{V_{j {\boxplus}_{|Z|} l}\} }\big)}}: l \in \{0,..,m_r+t_T\}\}$
    \EndFor
    \EndFor
    \EndFor
\EndProcedure
\end{algorithmic}
\end{algorithm}
The idea of the splitting of a subfile $W_{d_V,X_t,X_r}$ into packets $\{W_{d_V,X_t,X_r,Y}\}$ is that $Y$ denotes the receivers where the packet $W_{d_V,X_t,X_r,Y}$ is to be zero-forced. We now verify that the delivery scheme indeed delivers all the missing packets to all receivers. We do this in two steps.
\begin{itemize}
\item \textit{Verification of the Step 11 of Algorithm:} Note that the the packet $W_{d_{V_{j {\boxplus}_{|Z|} l}},X_t,{({S {\boxplus}_{|Z|} l}),Z\setminus \big({(S {\boxplus}_{|Z|} l ) \cup \{V_{j {\boxplus}_{|Z|} l}\} }\big)}}$ is cached at the $(m_r+1)$ of the $(m_r+t_T+1)$ receivers indexed by $Z$ given by the indices $S\boxplus_{|Z|}l.$ Thus Lemma \ref{ZF} can apply here, and the collection of packets in Step 11 can participate in a valid transmission.
\item \textit{Ensuring all missing packets are delivered:} Consider a demanded packet $W_{d_{V_a},X_t,X_r,Y}$. This packet is delivered in the unique transmission round consisting of the packets ${\cal P}(X_t,V_a,V_a\cup Y\cup X_r,X_r).$ Thus all missing packets of the demanded files are delivered to the corresponding receivers.
\end{itemize}

We now obtain the parameter $F_P$ which is the number of packets into which each demanded subfile is divided.
\begin{lemma}
\label{F_P expression}
$F_P=\frac{q^{\frac{(t_T+2m_r+2)(t_T-1)}{2}}}{(t_T-1)!}\prod\limits_{i=1}^{t_T-1}\gbinom{k_r-m_r-l_r-i}{1}$.
\end{lemma}

\begin{IEEEproof}
Consider an arbitrary demanded subfile $W_{d_{V_a},X_t,X_r}$ for some $X_r\in \mathbb{X}_r$. Finding $F_P$ is equivalent to counting the number of distinct sets $Y\in \mathbb{Y}$ such that $\{V_a\} \cup X_r \cup Y \in \mathbb{Z}$. This is equivalent to counting number distinct sets $ \{C_1,C_2,\cdots, C_{t_T-1}\}$ (such that $ C_i\in \mathbb{C}_r~\forall i\in [t_T-1]$ and $V_a+\sum\limits_{V_i\in X_r}V_i + \sum\limits_{i=1}^{t_T-1}C_i \in  \mathbb{E} $) which gives distinct $\{V_a\}\cup X_r \cup \{L_r\oplus C_1,L_r\oplus C_2, \cdots ,L_r\oplus C_{t_T-1}\} \in \mathbb{Z}$.
By following the similar proof technique, as employed in Lemma \ref{K_T,K_R,t_T,t_R,F expressions} we can write,

\begin{align*}
F_P&=\frac{\prod\limits_{i=0}^{t_T-2}(\theta(k_r)-\theta(m_r+l_r+1+i))}{(t_T-1)!~ q^{(t_T-1)(l_r-1)}}\\
&=\frac{\prod\limits_{i=1}^{t_T-1}(\theta(k_r)-\theta(m_r+l_r+i))}{(t_T-1)!~ q^{(t_T-1)(l_r-1)}}\\
&= \frac{1}{(t_T-1)!~ q^{(t_T-1)(l_r-1)}}\prod\limits_{i=1}^{t_T-1}\frac{q^{k_r}-q^{m_r+l_r+i}}{q-1}\\
&= \frac{q^{(m_r+l_r)(t_T-1)}\prod\limits_{i=1}^{t_T-1}q^{i}}{(t_T-1)!~ q^{(t_T-1)(l_r-1)}}\prod\limits_{i=1}^{t_T-1}\frac{q^{k_r-m_r-l_r-i}-1}{q-1}\\
&= \frac{q^{(m_r+1)(t_T-1)}q^{\frac{(t_T-1)(t_T)}{2}}}{(t_T-1)!}\prod\limits_{i=1}^{t_T-1}\gbinom{k_r-m_r-l_r-i}{1}\\
&= \frac{q^{\frac{(t_T+2m_r+2)(t_T-1)}{2}}}{(t_T-1)!}\prod\limits_{i=1}^{t_T-1}\gbinom{k_r-m_r-l_r-i}{1}.
\end{align*}
\end{IEEEproof}

\begin{remark}
The subpacketization of the proposed scheme is $F= F_C F_P=F_T F_R F_P$.
\end{remark}

We now summarize our scheme parameters in the following theorem. The proof of the theorem follows from the earlier lemmas and discussions in this section. The $\mathsf{DoF}$ follows from the observation that in Algorithm \ref{transmission algorithm}, each round serves $m_r+t_T+1$ packets to the same number of distinct receivers.

\begin{theorem}
Given a coded caching scheme as described in Section \ref{projective geometry based scheme}, with parameters $K_T,K_R,\frac{M_T}{N}=\frac{t_T}{K_T}, \frac{M_R}{N}=\frac{t_R}{K_R}, F_C,F_P$ (as per Lemma \ref{K_T,K_R,t_T,t_R,F expressions}, \ref{F_P expression}) we get a transmission scheme with subpacketization level $F=F_C F_P$  and $\mathsf{DoF}= m_r+t_T+1$.
\end{theorem}

Finally in Table \ref{table 1}, we give a numerical comparison of the schemes proposed in Sections \ref{modified1}, \ref{projective geometry based scheme} with the scheme given in \cite{hypercube_interference} (for the exact expressions for the parameters of \cite{hypercube_interference}, please refer to \cite{hypercube_interference}).
Note that we don't compare with \cite{interferencemanagement} because our scheme in Section \ref{modified1} achieves same $\mathsf{DoF}$ as \cite{interferencemanagement} with smaller subpacketization. The $-$ term indicates that there is no scheme in \cite{hypercube_interference} with the specific choice of parameters.
From the table it is clear that our scheme presented in Section \ref{projective geometry based scheme} outperforms the state of the art schemes in terms of the subpacketization with reduced $\mathsf{DoF}$.


\newcolumntype{g}{>{\columncolor{Gray}}c}


\begin{table}
\setlength{\tabcolsep}{4pt}
\renewcommand{\arraystretch}{1.5}
\centering
\resizebox{8.85cm}{!} {
\begin{tabular}{|c|c|c|c|c|c|c|c|c|}
\hline
$K_T$ & $K_R$ & $\frac{M_T}{N}$ & $\frac{M_R}{N}$ & DoF & DoF  & $F$ &  $F$  & $F$ \\

& & & & \cite{hypercube_interference}, \ref{modified1}  & \ref{projective geometry based scheme} & \cite{hypercube_interference}& \ref{modified1} & \ref{projective geometry based scheme}  \\

\hline
7 & 31 & .428 & .097 & 6 & 5 & $3\times 10^{6}$ & $10^{7}$ &
$2\times 10^{6}$\\
\hline
7 & 63 & .428 & .111 & 10 & 6 & - & $10^{13}$ & $7\times 10^{8}$
\\
\hline
7 & 127 & .428 & .055 & 10 & 6 & - & $10^{16}$ & $4\times 10^{10}$
\\
\hline

13 & 364 & .308 & .011 & 8 & 6 & $3\times 10^{16}$ & $ 10^{18}$ &
$2\times 10^{13}$
\\
\hline

40 & 364 & .10 & .011 & 8 & 6 & $3\times 10^{18}$ & $10^{20}$ &
$3\times 10^{14}$
\\
\hline

40 & 1093 & .10 & .004 & 8 & 6 & $7\times 10^{21}$ & $ 10^{24}$ &
$8\times 10^{16}$
\\
\hline

\end{tabular}
}
\caption{For some specific values of $K_T,K_R,\frac{M_T}{N}, \frac{M_R}{N}$, we compare the results of \cite{hypercube_interference} with the schemes presented in Sections \ref{modified1}, \ref{projective geometry based scheme} }
\label{table 1}
\end{table}


\section{Analysis of the scheme}
\label{analysis}

In this section, we analyse the asymptotic behaviour of subpacketization$(F)$ and $\mathsf{DoF}$ of the scheme presented in Section \ref{projective geometry based scheme} for large $K_R$ (this represents a practical regime of interest) by upper bounding $\frac{M_T}{N}, \frac{M_R}{N}$ by some constant.
We use the following lemma.

\begin{lemma}\cite{PK}
\label{approximations}
Let $a,b,f \in \mathbb{Z}^+$ and $q$ is some prime power. Then,
\begin{align}
\label{eqn qbinom inequality 1}
&q^{(a-b)b}&\leq &\gbinom{a}{b} \leq & q^{(a-b+1)b}\\
\label{eqn qbinom inequality 2}
&q^{(a-f-1)b} &\leq &\frac{\gbinom{a}{b}}{\gbinom{f}{b}}\leq & q^{(a-f+1)b}
\end{align}
\end{lemma}

From Lemma \ref{K_T,K_R,t_T,t_R,F expressions} we have,
$K_T = \gbinom{k_t-l_t+1}{1}$ and $K_R = \gbinom{k_r-l_r+1}{1}$.
We first upper bound $\frac{M_T}{N}$ and $\frac{M_R}{N}$ by some constants. Consider,
\[\frac{M_T}{N} =\frac{t_T}{K_T}=\frac{\gbinom{m_t+1}{1}}{\gbinom{k_t-l_t+1}{1}}\stackrel{(\ref{eqn qbinom inequality 2})}{\leq} q^{m_t-k_t+l_t+1}.\]

To upper bound $\frac{M_T}{N}$ by a constant, let $k_t-m_t-l_t=\alpha$, for some constant $\alpha \in \mathbb{Z}^{+}$. We thus have $\frac{M_T}{N} \leq \frac{1}{q^{\alpha-1}}$.

Similarly to upper bound $\frac{M_R}{N}$ by a constant, let $k_r-m_r-l_r=\beta$, for some constant $\beta \in \mathbb{Z}^{+}$.
We thus have $\frac{M_R}{N} \leq \frac{1}{q^{\beta-1}}$.

Now by using $(\ref{eqn qbinom inequality 1})$ we have, 
$q^{k_t-l_t} \leq K_T \leq q^{k_t-l_t+1}$.

So we have,
\begin{align}
    \label{eqn inequality k_t-l_t}
    {\log_{q}{K_T}}-1 \leq (k_t-l_t)\leq {\log_{q}{K_T}}
\end{align}
Similarly,
\begin{align}
    \label{eqn inequality k_r-l_r}
    {\log_{q}{K_R}}-1 \leq (k_r-l_r)\leq {\log_{q}{K_R}}
\end{align}

The $F_T,K_T$ expressions are same that of $F,K$ in \cite{haribhavanaprasad} respectively. Therefore from \cite{haribhavanaprasad} (Section V) we have $F_T=q^{O\left((\log_{q}{K_T})^2\right)}$. Similarly $F_R=q^{O\left((\log_{q}{K_R})^2\right)}$.

Now we will analyze the asymptotics of $F_P$.
From Lemma \ref{F_P expression} we have,

\begin{align*}
F_P &= \frac{q^{(m_r+1)(t_T-1)}\left(\prod\limits_{i=1}^{t_T-1}q^{i}\right)}{(t_T-1)!}\prod\limits_{i=1}^{t_T-1}\gbinom{k_r-m_r-l_r-i}{1} \\
& \stackrel{(\ref{eqn qbinom inequality 1})}{\leq} \frac{q^{(m_r+1)(t_T-1)}\left(\prod\limits_{i=1}^{t_T-1}q^{i}\right)}{(t_T-1)!}\prod\limits_{i=1}^{t_T-1} q^{k_r-m_r-l_r-i} \\
&= \frac{q^{(m_r+1)(t_T-1)}q^{(k_r-m_r-l_r)(t_T-1)}}{(t_T-1)!}.
\end{align*}
Hence,
\begin{align}
\label{eqn F_P ineuqlity}
F_P &\leq \frac{q^{(k_r-l_r+1)(t_T-1)}}{(t_T-1)!}.
\end{align}

From Lemma \ref{K_T,K_R,t_T,t_R,F expressions} we have, 
$t_T = \gbinom{m_t+1}{1}$. Now we can write,
$q^{m_t}\stackrel{(\ref{eqn qbinom inequality 1})}{\leq} t_T$. 
From this we can write, $\frac{1}{(t_T-1)!}\leq \frac{1}{(q^{m_t}-1)!}$.

Also from (\ref{eqn inequality k_t-l_t}) we have, 

${\log_{q}{K_T}}-1 \leq m_t+\alpha$.
which can be written as 
$q^{{\log_{q}{K_T}}-1-\alpha}-1 \leq q^{m_t}-1$.

Therefore we have,
\begin{align*}
\frac{1}{(t_T-1)!} \leq \frac{1}{( q^{m_t}-1)!}\leq \frac{1}{(q^{{\log_{q}{K_T}}-1-\alpha}-1)!}.
\end{align*}

Consider,
$t_T -1 \stackrel{(\ref{eqn qbinom inequality 1})}{\leq}  q^{m_t+1}-1$.
But
$m_t \stackrel{(\ref{eqn inequality k_t-l_t})}{\leq} {\log_{q}{K_T}} -\alpha$.

Therefore we have,
$t_T -1 \leq q^{{\log_{q}{K_T}} -\alpha +1} -1$.

Also we have ,
\begin{align*}
(k_r-l_r+1) \stackrel{(\ref{eqn inequality k_r-l_r})}{\leq} {\log_{q}{K_R}}+1.    
\end{align*}

By using these inequalities in (\ref{eqn F_P ineuqlity}) we have,

\[F_P\leq \frac{q^{({\log_{q}{K_R}}+1)(q^{{\log_{q}{K_T}} -\alpha +1} -1)}}{(q^{{\log_{q}{K_T}}-1-\alpha}-1)!}.\]

Now by using Stirling's approximation and simple manipulations we have,
$F_P=q^{O\left(K_T\log_{q}{K_R}\right)}$.

Therefore,

\begin{align*}
F&=F_T F_R F_P \\ &=q^{O\left((\log_{q}{K_T})^2\right)}q^{O\left((\log_{q}{K_R})^2\right)}q^{O\left(K_T\log_{q}{K_R}\right)} \\
&=q^{O\left(K_T+(\log_{q}{K_R})^2\right)}.    
\end{align*}

Now we will analyse the asymptotics of $\mathsf{DoF}=m_r+t_T+1$ .
From \ref{eqn inequality k_r-l_r} we have 
\begin{align}
\label{eqn m_r+1 ineuqality}
{\log_{q}{K_R}}-\beta \leq (m_r+1)\leq {\log_{q}{K_R}}-\beta +1    
\end{align}

By using \ref{eqn qbinom inequality 1} we have, 
$q^{m_t}\leq t_T \leq q^{m_t+1}$.

By using \ref{eqn inequality k_t-l_t} we have, \[{\log_{q}{K_T}}-1-\alpha \leq m_t \leq {\log_{q}{K_T}}-\alpha.\]
From this we can write

$q^{{\log_{q}{K_T}}-1-\alpha}\leq q^{m_t}$ and $q^{m_t+1} \leq q^{{\log_{q}{K_T}}+1-\alpha}$.

So we have, 
\begin{align}
\label{eqn t_T inequality}
q^{{\log_{q}{K_T}}-1-\alpha}\leq t_T \leq q^{{\log_{q}{K_T}}+1-\alpha}.    
\end{align}

Hence by using (\ref{eqn m_r+1 ineuqality}) and (\ref{eqn t_T inequality}) we have

${\log_{q}{K_R}}-\beta +q^{{\log_{q}{K_T}}-1-\alpha}\leq m_r+1+t_T \leq  {\log_{q}{K_R}}-\beta +1+q^{{\log_{q}{K_T}}+1-\alpha}.$

Therefore
$\mathsf{DoF} = \Theta(log_qK_R+K_T)$.



\bibliographystyle{IEEEtran}
\bibliography{IEEEabrv,cite.bbl}

\end{document}